\documentclass[journal=jpclcd,manuscript=letter]{achemso}
\usepackage[version=3]{mhchem} 
\usepackage{color}
\usepackage{bm}

\newcommand{\Fig}[1]{Figure~\ref{#1}}

\newcommand{\Figs}[1]{Figures~\ref{#1}}

\newcommand{\Equation}[1]{Equation~\ref{#1}}
\newcommand{\Equations}[1]{Equations~\ref{#1}}
\author{Jiaan~Cao}
\affiliation{Hefei National Research Center for Physical Sciences at the Microscale, 
University of Science and Technology of China, Hefei, Anhui 230026, China}
\altaffiliation{These authors contributed equally to this work.}
\author{Lyuzhou~Ye}
\email{lzye@ustc.edu.cn}
\affiliation{Hefei National Research Center for Physical Sciences at the Microscale, 
University of Science and Technology of China, Hefei, Anhui 230026, China}
\altaffiliation{These authors contributed equally to this work.}
\author{Dawei~He}
\affiliation{Hefei National Research Center for Physical Sciences at the Microscale, 
University of Science and Technology of China, Hefei, Anhui 230026, China}

\author{Xiao~Zheng}
\email{xzheng@fudan.edu.cn}
\affiliation{Department of Chemistry, Fudan University, Shanghai 200433, China}

\author{Shaul~Mukamel}
\affiliation{Department of Chemistry and Department of Physics and Astronomy,
University of California, Irvine, CA 92697, USA}

\title{Magnet-Free Time-Resolved Magnetic Circular Dichroism with Pulsed Vector Beams}

\begin{document}

\begin{tocentry}
%
%
%
%
%
%
\includegraphics[width=\columnwidth]{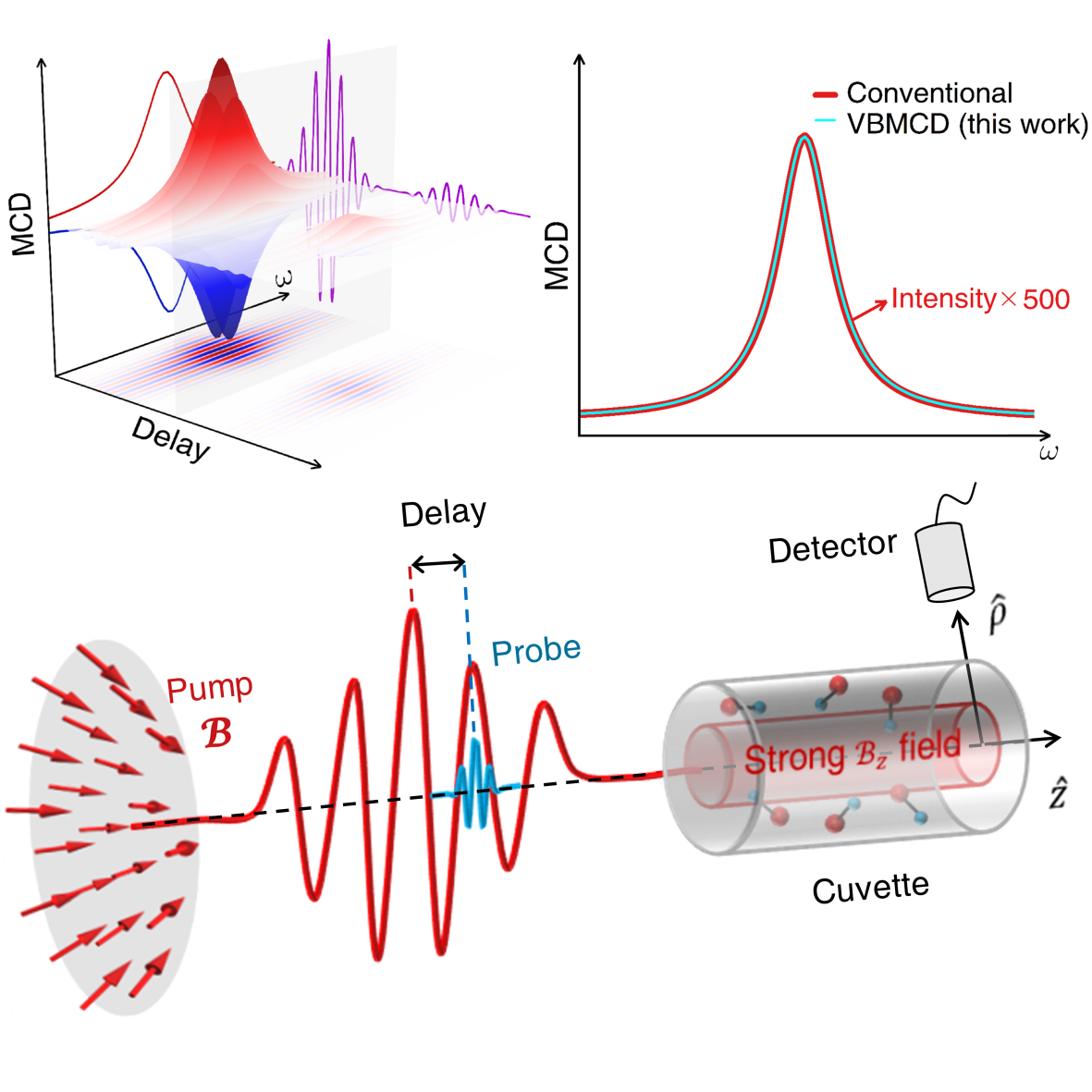}
\end{tocentry}

\begin{abstract}
Magnetic circular dichroism (MCD) is a widely used spectroscopic technique which reveals valuable information about molecular geometry and electronic structure. However, the weak signal and the necessary strong magnets impose major limitations on its application. We propose a novel protocol to overcome these limitations by using pulsed vector beams (VBs), which consist of nanosecond gigahertz pump and femtosecond UV-Vis probe pulses. By virtue of the strong longitudinal electromagnetic fields, the MCD signal detected by using the pulsed VBs is greatly enhanced compared to conventional MCD performed with plane waves. Furthermore, varying the pump-probe time delay allows to monitor the ultrafast variation of molecular properties.
\end{abstract}


Over the past decades,
magnetic molecules, ranging from simple radicals to single molecule magnets,
have opened up potential applications in many areas, such as molecular spintronics,\cite{serrano2020quantum,coronado2020molecular,oh2021scalable}
magnetic refrigeration,\cite{zheng2012co,richmond2012flow,karotsis2010mniii4lniii4,liu2019situ}
and electrocatalysis.\cite{chalkley2018fe,nesbit2019characterization}
Various optical spectroscopic techniques have been developed to investigate
the electronic and magnetic properties of molecules and understand the magneto-structural correlation,
and magnetic circular dichroism (MCD) is one of the most widely used tools.\cite{mason2007magnetic,gonidec2010probing,toriumi2015plane}
In this technique,
a strong static magnetic field is applied to induce Zeeman splitting in molecules
by lifting the degeneracy of the electronic states.\cite{han2020magnetic,kjaergaard2012ab}
The MCD signal is then given by
the differential absorption of the left- and right-circularly polarized (LCP and RCP) light
passing through the molecular sample.\cite{mason2007magnetic}
MCD has demonstrated its superiority over linear absorption spectroscopy
in determining electronic transitions,\cite{han2020magnetic,yao2012electronic,daumann2015new}
as well as in revealing information about the degeneracy and symmetry of electronic states.\cite{mason2007magnetic,toriumi2015plane,kobayashi2011circular}

Despite its rapid progress and wide applications, MCD still faces several challenges:
First, being the differential absorption of the LCP and RCP plane waves (PWs),
the intensity of conventional MCD signal is generally 2 to 3 orders of magnitude weaker than the absorption signal.\cite{heit2019magnetic,guclu2016photoinduced,mitic2003rapid,snyder2013circular}
Second, a strong static magnetic field is needed to induce the Zeeman splitting in conventional MCD experiments. Although numerous types of magnets have been designed,\cite{mason2007magnetic,han2020magnetic,daifuku2014combined}
their practical use can be cumbersome, because of the required large-sized devices and cryogenic temperatures.\cite{mason2007magnetic,daifuku2014combined,tomita2003high,durrell2014trapped}
The first challenge concerning weak signal can be overcome by replacing the LCP and RCP PWs
by left- and right-circularly polarized beams with nonzero longitudinal components (LCPL and RCPL),\cite{ye2021enhancing}
which are realized by superposing the azimuthally and radially polarized (AP and RP) vector beams (VBs).
It has been proposed that VBs could greatly enhance the circular dichroism (CD) signals of chiral molecules,\cite{ye2021enhancing}
thanks to the strong electromagnetic fields generated by the VBs.
However, it is difficult to probe time-varying molecular properties with continuous-wave VBs,
which raises a third challenge for MCD measurements.

Inspired by the recent progress in the design and implementation of VBs\cite{guclu2016photoinduced,zhan2009cylindrical,blanco2018ultraintense,levy2019mathematics}
and gigahertz (GHz) spectroscopy,\cite{rostov2016superradiant,mesyats2017phase}
in this work we propose a novel protocol for magnet-free time-resolved MCD measurements,
which utilizes pulsed VBs and will be denoted VBMCD.

%

VBs have cylindrical symmetry along the beam axis.
In the cylindrical coordinate system, the electric and magnetic fields of
a pulsed AP VB are expressed as
\begin{align}
\bm{\mathcal E}_{\rm A}(\bm r,t) &= \hat{\phi}\, {\mathcal H}(\bm r) \, e^{i\left(Kz-\Omega t\right)} \, \mathcal G(t), \\
\bm {\mathcal B}_{\rm A}(\bm{r},t) &= \left[ \hat{\rho} \left(-\frac{K}{\Omega} \right) + 
\hat{z} \left( -\frac{i}{\Omega}\right) 
\left( \frac{1}{\rho} + \frac{\partial}{\partial \rho} \right)\right]
{\mathcal H} (\bm r)\, e^{i\left(Kz-\Omega t\right)}\, \mathcal G(t). 
\end{align}
Here, $\hat{\rho}$, $\hat{\phi}$, and $\hat{z}$ represent the radial, azimuthal, and longitudinal unit vectors, respectively;
$\bm r \equiv (\rho,\phi,z)$, and $\Omega$ is the central frequency.
The spatial profile assumes a
Hermite-Gaussian form, \cite{zhan2009cylindrical}
\begin{equation}
\mathcal H(\bm r) =  \mathcal A_0\frac{4\rho}{\sqrt{\pi} [W(z)]^2}\, {\rm exp}\!\left[-\frac{\rho^2}{[W(z)]^2}\right]
{\rm exp}\!\left[{ \frac{iK\rho^2}{2R (z)}}\right] {\rm exp}\!\left[{-2i\arctan\!\left(\frac{z}{z_R}\right)}\right], \label{eqS1}
\end{equation}
where $\mathcal A_0$ is the amplitude, $K$ is the wavenumber, $W_0$ is the beam waist, $z_{R}=\frac{1}{2}K W_{0}^2$ is the Rayleigh length, 
$W(z)=W_{0}\sqrt{1+z^2/z_{R}^2}$ is the beam width, and $R(z) = z + z_{R}^2/z$ is the radius of curvature.
The temporal profile assumes a Gaussian form,
\begin{equation}
\mathcal G(t)= {\rm exp}\!\left[-4{\rm ln 2}\frac{(t-T_c)^2}{\tau_D^2}\right], 
\end{equation}
with $T_c$ and $\tau_D$ being the central time and  the duration time, respectively.
%
%
%
The corresponding electric and magnetic fields of the pulsed RP VB are respectively \cite{zhan2009cylindrical}
\begin{align}
\bm{\mathcal{E}}_{\rm R}(\bm r,t) &= \left[\hat{\rho}
+ \hat{z} \left( \frac{i}{K}\right)
\left(\frac{1}{\rho} + \frac{\partial}{\partial \rho} \right) \right]{\mathcal H} (\bm r)\, e^{i\left(Kz-\Omega t\right)}\, \mathcal G(t), \\
\bm{\mathcal{B}}_{\rm R}(\bm r,t) &=  \hat{\phi}\left( \frac{K}{\Omega}\right){\mathcal H}(\bm r) \, e^{i\left(Kz-\Omega t\right)} \, \mathcal G(t).
\end{align}

\begin{figure}[h]
\centering
  \includegraphics[width=0.9\columnwidth]{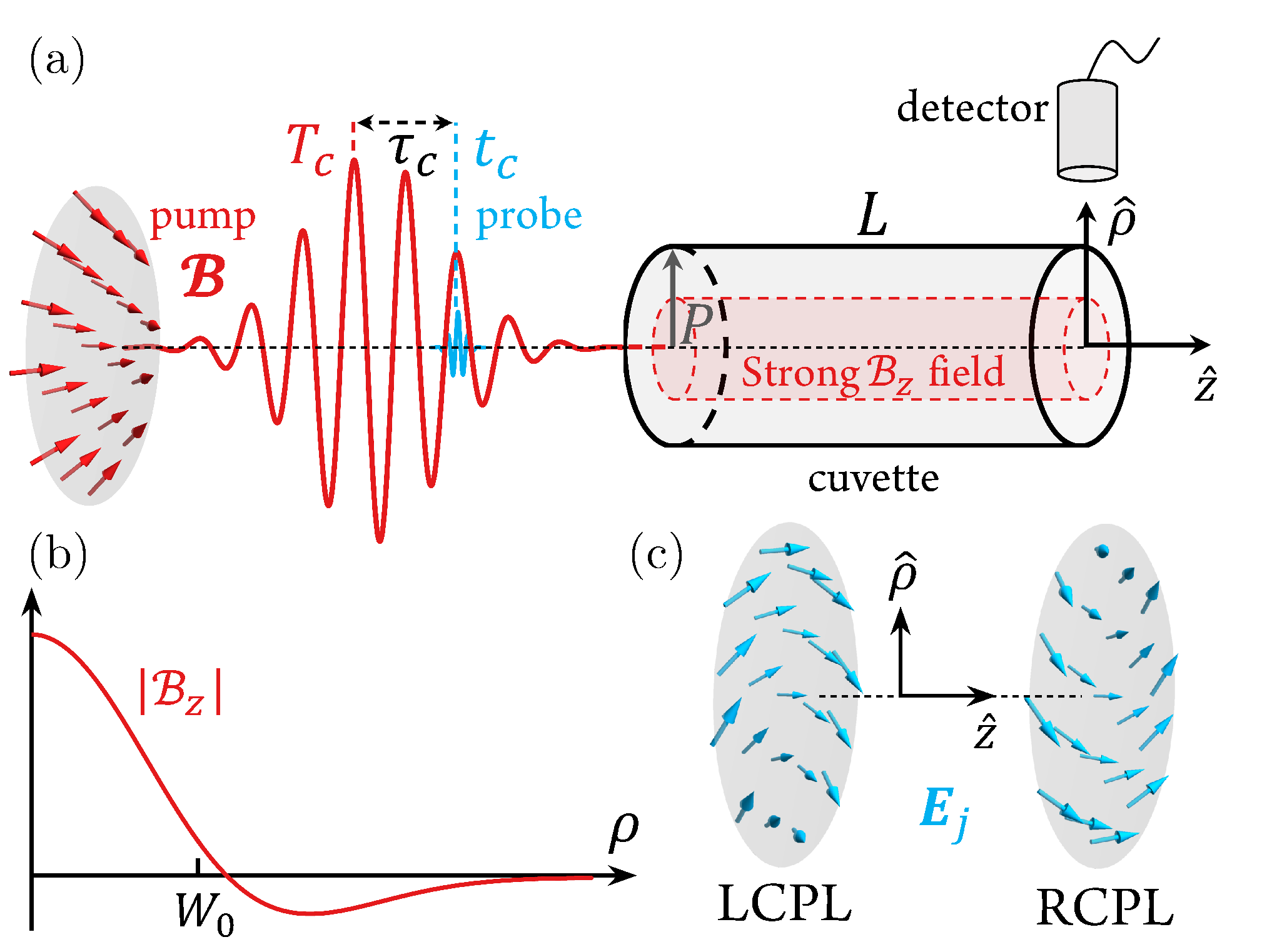}
  \caption{(a) Schematic of the proposed VBMCD protocol. The red cylinder represents the paraxial region where a strong localized magnetic field generated by the pump VB exists. ${\hat \rho}$ and ${\hat z}$ denote the radial and longitudinal directions, and $T_c$ and $t_c$ are the centers of pump and probe pulses, respectively. (b) Variation of $|\mathcal B_z|$ with respect to $\rho$, where $W_0$ is the pump beam waist. (c) Vector field plots of the LCPL and RCPL probe VBs.}  \label{fig1}
\end{figure}

In this work the $0.1\sim 10$~T static magnetic field typically used in conventional MCD is replaced by a temporally- and spatially-varying magnetic field $\bm{\mathcal{B}}(\bm r, t)=\bm{\mathcal{B}}_{\rm A}(\bm r, t)$, which is generated by an AP VB (\Fig{fig1}a).
The longitudinal component of $\bm{\mathcal{B}}_{\rm A}$, denoted as $\mathcal B_z$, is tightly localized around the beam axis, and its amplitude decreases rapidly with the radial distance $\rho$ (\Fig{fig1}b).\cite{zhan2009cylindrical,blanco2018ultraintense,levy2019mathematics}
To create an appreciable Zeeman splitting in the molecule, a pulsed VB is utilized as the pump beam to supply a strong 
longitudinal field $\mathcal B_z$.
Two pulsed probe beams, LCPL and RCPL\cite{ye2021enhancing} (\Fig{fig1}c), which are collinear with the pump VB, 
are employed to measure the MCD signals.
The electric and magnetic fields of the probe pulses are explicitly
expressed as\cite{ye2021enhancing}
\begin{align}
\left(
  \begin{array}{cc}
    {{\bm E}}_{j}(\bm r,t) \\
    {{\bm B}}_{j}(\bm r,t) \\
  \end{array}
\right)
= \frac{-1}{\sqrt{2}}
\left(
  \begin{array}{cc}
    {{\bm E}}_{\rm A}(\bm r,t) & {{\bm E}}_{\rm R}(\bm r,t)  \\
    {{\bm B}}_{\rm A}(\bm r,t) & {{\bm B}}_{\rm R}(\bm r,t)  \\
  \end{array}
\right)
\left(
  \begin{array}{cc}
    c_j \\
    i   \\
  \end{array}
\right)   ,  \label{eq_1}
\end{align}
where $c_j= 1$ for $j = {\rm LCPL}$ and $-1$ for $j ={\rm RCPL}$,
and $\bm{E}_{\rm A/R}$ and $\bm{B}_{\rm A/R}$ denote the electric
and magnetic fields of the AP/RP VBs for designing the LCPL and RCPL, respectively.
The spatial and temporal profiles of the probe VBs, denoted as $H(\bm r)$ and $G(t)$, 
have the same forms as those of the pump, but the values of the involving parameters are different. These include amplitude $A_0$,
wavenumber $k_j$, beam waist $w_0$, pulse central time $t_c$ and duration time $\tau_d$.

To ensure that the Zeeman states are detectable by the probe VBs,
the strength of localized magnetic field should reach $0.1\sim 10$~T.\cite{han2020magnetic}
Moreover, to acquire nonzero time-averaged MCD signal with the oscillatory pump pulse,
the duration of the probe pulses should be much shorter than that of the pump pulse, so that the longitudinal magnetic field $\mathcal B_z$ of the pump beam is almost unchanged during the action of the probe pulse. These suggest the use of a nanosecond GHz pump pulse and femtosecond UV-Vis probe pulses.
Generation of such pulsed VBs is well within the capabilities of current technology.\cite{levy2019mathematics,rostov2016superradiant,mesyats2017phase}
Furthermore, the time delay between the pump and probe pulses can be precisely tuned,
thus allowing for the detection of time-varying molecular properties. 
%
%

%


%
In the following, we compare the VBMCD signal to conventional MCD.
The light-matter interaction Hamiltonian for our setup is $H_{\rm int}=
-\hat{\bm \mu}\cdot\bm{E}_j-\hat{\bm m} \cdot \bm{\mathcal B}-\frac{1}{2}\hat {\bm Q}:\nabla\bm{E}_j+{\rm H.c.}$, where $\hat{\bm \mu}$, $\hat{\bm m}$, and $\hat {\bm Q}$ are the electric transition dipole, magnetic transition dipole, and electric transition quadrupole operators, respectively. 
The contribution of the electric quadrupole term to the MCD signal vanishes when performing rotational averaging over randomly oriented molecules in solutions or in the gas phase.\cite{mason2007magnetic,cho_two-dimensional_2009}
The interactions between the molecular transition dipoles and the electric component of the pump VB and the magnetic component of the probe pulses are neglected,
because the former is off-resonant with the electronic excitations in molecules due to much lower frequency of the pump pulse, while the latter is several orders of magnitude weaker than $-\hat{\bm \mu}\cdot \bm E_j$.\cite{mason2007magnetic}
The wavenumber $k_j$ of the LCPL and RCPL pulses can be obtained by first-order time-dependent perturbation theory;\cite{mukamel1999principles} see Section~S1 in the Supporting Information (SI). Their difference, $\Delta k\equiv k_{\rm LCPL}-k_{\rm RCPL}$, which largely determines the lineshape of the MCD spectrum, is expressed as the sum of three Faraday terms, denoted by $A(\omega)$, $B(\omega)$ and $C(\omega)$, as follows,
\begin{align}
\Delta k(\omega)
&= \frac{n \mu_0 c }{3\hbar^2} \,\mathcal{B}_z\left(\bm r,t\right) \left[A(\omega) + B(\omega) + C(\omega) \right] , \label{eq_2} \\
\Lambda(\omega) &= \tilde \Lambda(\omega,\Omega)+\frac{{\mathcal B}^\ast_{z}\left(\bm r,t\right)}{{\mathcal B}_{z}\left(\bm r,t\right)}\,\tilde \Lambda(\omega,-\Omega), \quad \Lambda \in \{A, B, C\},  \\
\tilde A (\omega;\Omega ) &= \frac{i\left(\omega+\Omega \right)^2}{2N_g \,\omega} \sum_{g_\alpha,l_\beta} \sum_{g_{\alpha^\prime},l_{\beta^\prime}}
\frac{\partial f_{l_\beta g_\alpha}(\omega)}{\partial \omega}\nonumber\\
& \quad \left(\bm{m}_{l_{\beta} l_{\beta^\prime}}\delta_{g_\alpha g_{\alpha^{\prime}}}-\bm{m}_{g_\alpha g_{\alpha^{\prime}}}\delta_{l_{\beta} l_{\beta^\prime}}\right)\cdot \left(\bm{\mu}_{g_{\alpha^{\prime}} l_{\beta^{\prime}}} \times \bm{\mu}_{l_\beta g_\alpha}\right), \label{eq_3}  \\
\tilde B (\omega;\Omega ) &= \frac{-i\left(\omega+\Omega \right)^2}{2N_g \,\omega} \sum_{g_\alpha, l_\beta}\left[f_{l_\beta g_\alpha}(\omega)+f_{l_\beta g_\alpha}(\omega+\Omega)\right]\nonumber\\
& \quad \left[\sum_{v_\gamma, v_\gamma \neq g_\alpha} \frac{\left(\bm{\mu}_{l_\beta g_\alpha}\times \bm{\mu}_{v_\gamma l_\beta}\right)\cdot \bm{m}_{g_\alpha v_\gamma}}{\Omega+\omega_{v_\gamma g_\alpha}}
+\sum_{v_\gamma, v_\gamma \neq l_\beta} \frac{\left(\bm{\mu}_{g_\alpha v_\gamma}\times \bm{\mu}_{l_\beta g_\alpha}\right)\cdot
\bm{m}_{v_\gamma l_\beta}}{\Omega+\omega_{l_\beta v_\gamma}}\right], \label{eq_4} \\
\tilde C(\omega;\Omega) &=\frac{-i\left(\omega+\Omega \right)^2}{2N_g \, \omega}\frac{\hbar}{k_B T} \sum_{g_\alpha,g_{\alpha^\prime},l_\beta}f_{l_\beta g_\alpha}(\omega+\Omega)
\left(\bm{\mu}_{l_\beta g_\alpha}\times \bm{\mu}_{g_{\alpha^\prime} l_\beta}\right)\cdot \bm{m}_{g_\alpha g_{\alpha^\prime}} .  \label{eq_5}
\end{align}
Here, $n$ is the molecular concentration,
$\mu_0$ is the magnetic permeability,
$c$ is the speed of light,
and $f_{l_\beta g_\alpha}(\omega) \equiv (\omega - \omega_{l_\beta g_\alpha} + i\eta)^{-1}$,
with $g_\alpha$ and $l_\beta$ labeling the states in the degenerate sublevels of
the ground $(g)$ and excited $(l)$ states, respectively.
$T$ is the temperature,
$N_g$ is the degeneracy of the ground state manifold,
$\bm{\mu}_{l_\beta g_\alpha}$ and $\bm{m}_{l_\beta g_\alpha}$
denote the electric and magnetic transition dipole moments
between states $l_\beta$ and $g_\alpha$, respectively.
The expressions in Equations \eqref{eq_3}--\eqref{eq_5} reduce to their counterparts for the conventional MCD 
when the pump central frequency $\Omega=0$.
Therefore, the conventional MCD spectrum measured with PW beams
can be recovered by replacing $\mathcal B_z$ with a static magnetic field $B_z$ (Section~S1 in the SI).
The Faraday $A$, $B$ and $C$ terms in Equation~\eqref{eq_2} are functions of the probe central frequency $\omega$. They provide valuable information about the degeneracy and symmetry of electronic states.
Specifically, the $A$ term arises from the Zeeman splitting of orbitally degenerate excited states.
It has a characteristic derivative lineshape and is independent of temperature.
The $B$ term originates from the magnetic-field-induced mixing of the zero-field nondegenerate states. 
It is also temperature-independent, but usually much smaller compared to the other two terms.
The $C$ term exhibits a strong inverse-temperature dependence. It reveals the information about the ground state population 
in the presence of the Zeeman splitting due to the external magnetic field, 
and is nonvanishing only for molecules with degenerate ground states. \cite{mason2007magnetic,kjaergaard2012ab,han2020magnetic}

%

For a PW beam, the longitudinal components of electric and magnetic fields are absent, 
and thus the luminous flux (magnitude of Poynting vector) is entirely along the axial direction of beam. 
Consequently, the flux can be gathered only in the longitudinal direction for the conventional MCD. 
In contrast, VBs have strong longitudinal components of electric and magnetic fields, 
and the luminous flux is nonzero in the radial direction. 
%
By using a cylindrical sample cuvette aligned coaxially with the probe pulses
(\Fig{fig1}a), the luminous flux in the radial direction, 
gathered by the detector placed at a given azimuthal angle $\phi$,  
is $I_{\rho}(\omega,\tau_c;k_j) =\frac{1}{\mu_0}\int dt\,\int_{0}^{P} \rho {\bm E}_{j}
\times {\bm B}_{j}^\ast \cdot d{\bm \rho} + {\rm c.c.}$,
with $P$ being the radius of interaction cross section between the probe beams and the molecular sample.
Because of their cylindrical symmetry, both ${\bm E}_{j}$ and ${\bm B}_{j}$ are independent of $\phi$, 
and hence the luminous flux $I_{\rho}(\omega,\tau_c;k_j)$ detected at any $\phi$ is identical. 
Therefore, using a ring-like detector encircling the axis of cuvette is most favourable 
for the collection of luminous flux.

The linear absorption spectrum is 
$\epsilon_\rho(\omega,\tau_c;k_j) =
(n_0L)^{-1} {\rm ln}[I_{\rho}(\omega,\tau_c;k_0)/I_{\rho}(\omega,\tau_c;k_j)]$,
where $\tau_c$ is the time delay between the pump and probe pulses,
$n_0$ is a reference molecular concentration,
$L$ is the length of the sample cuvette,
and $k_0=\omega/c$ is the wavenumber of the probe VBs in vacuum.
The resulting VBMCD spectrum is explicitly expressed as
\begin{align}
\Delta \epsilon_\rho(\omega, \tau_c) &\equiv
\epsilon_\rho(\omega, \tau_c; k_{\rm LCPL})- \epsilon_\rho(\omega, \tau_c; k_{\rm RCPL})  \nonumber \\
& = \frac{1}{n_0L}\, {\rm ln}\!\left[\frac{I_{\rho}(\omega,\tau_c;k_{\rm RCPL})}{I_{\rho}(\omega,\tau_c;k_{\rm LCPL})}\right].  \label{eq_6}
\end{align}
Clearly, the VBMCD spectroscopic signal characterizes the difference between
the linear absorption spectra measured by the LCPL and RCPL beams, 
rather than their absolute amplitudes. 
Thus, the intensity of the VBMCD signal should be distinguished from 
that of luminous flux collected by the detector. 
Particularly, although the luminous flux in the radial direction is 
relatively weaker than in the longitudinal direction, 
it is sufficiently bright to be harvested by modern detectors. 
Therefore, the VBMCD signal can be significantly enhanced over that of the conventional MCD, 
even with a relatively weaker luminous flux received
by the detector.

\begin{figure}[h]
\centering
  \includegraphics[width=0.9\columnwidth]{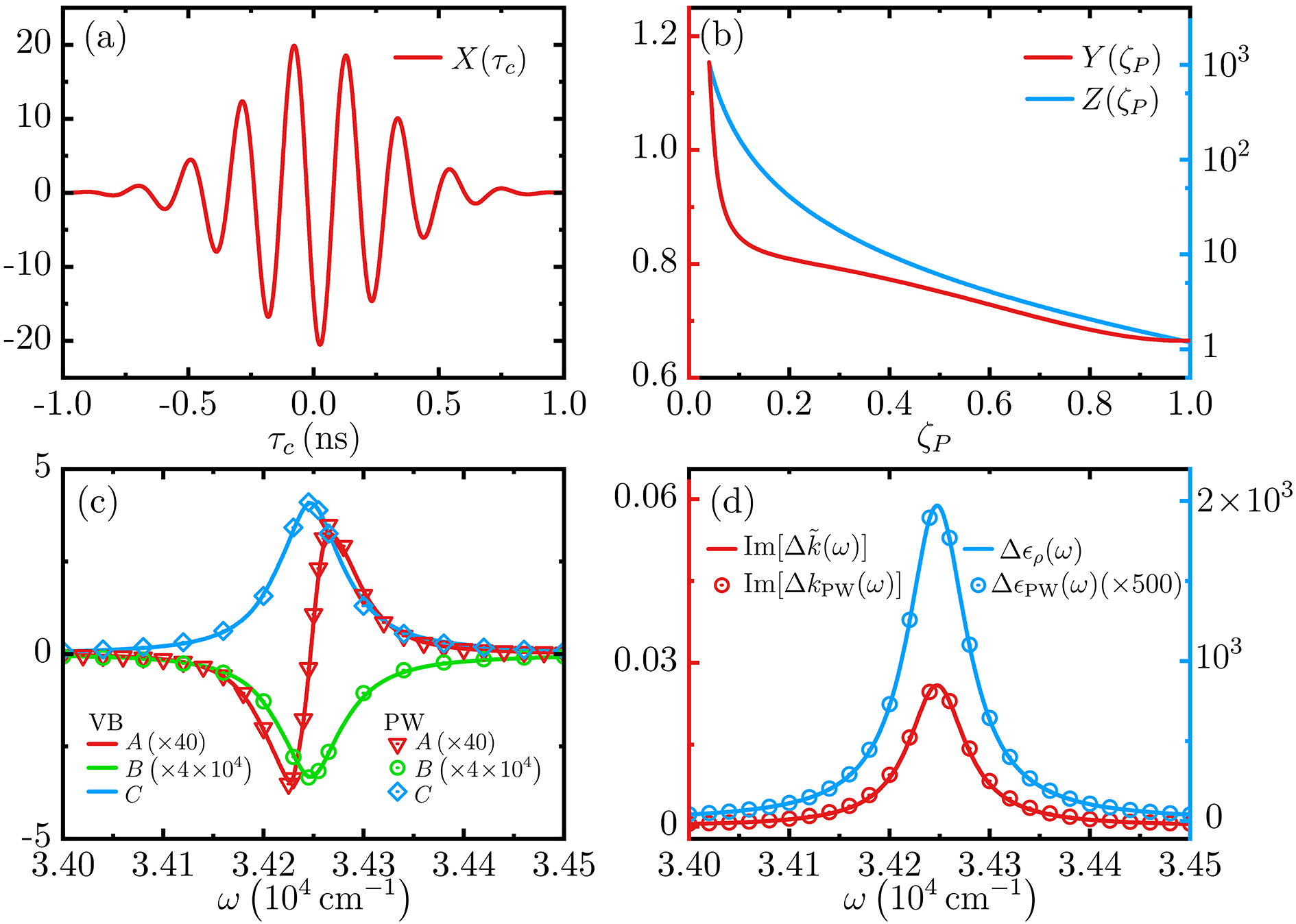}
  \caption{Characteristic functions which determine the intensity of the VBMCD spectrum: 
  (a) $X(\tau_c)$ and(b) $Y(\zeta_P)$ and $Z(\zeta_P)$. (c) The Faraday $A$, $B$ and $C$ terms of the VBMCD and the conventional MCD with PWs for the OH radical. (d) ${\rm Im}[\Delta \tilde k(\omega)]$ versus ${\rm Im}[\Delta k_{\rm PW}(\omega)]$ (left axis), and $\Delta \varepsilon_\rho(\omega)$ versus $\Delta \epsilon_{\rm PW}(\omega)$ (right axis) for the OH radical. For a direct comparison, $\Delta \epsilon_{\rm PW}(\omega)$ is amplified manually by 500 times. The parameters adopted for the simulations are listed in Sections~S2 and S3 in the SI. 
The molecular properties are obtained from quantum chemistry calculations\cite{lee1988development,becke98density,hehre1972self,klene2003new,dunning1989gaussian} 
by using the ORCA\cite{neese2020orca} and the MOLCAS\cite{Aquilante2016Feb} program packages.}
  \label{fig2}
\end{figure}

To elucidate the origin of signal enhancement, we express the VBMCD spectrum $\Delta \epsilon_\rho$ as
\begin{align}
\Delta \epsilon_\rho(\omega, \tau_c)
& \simeq \frac{{\rm Im}[\Delta \tilde{k}(\omega)]}{n_0}
\frac{X(\tau_c)Y(\zeta_P)Z(\zeta_P)}{2\zeta_L^2}. \label{eq_7}
\end{align}
Here, $\zeta_L\equiv L/w_0$ and $\zeta_P\equiv P/w_0$ are the reduced path length and reduced radius of the cuvette, respectively, with $w_0$ being the probe beam waist. $\Delta \tilde k(\omega)$ is $\Delta k(\omega)$ evaluated at $\tau_c=0$ and $\rho=W_0$. It dominates the lineshape of the VBMCD spectrum, which resembles closely the conventional MCD spectrum. $X(\tau_c)$, $Y(\zeta_P)$ and $Z(\zeta_P)$ are three characteristic functions which determine the intensity of the measured spectrum (see Section~S1 in the SI for more details). They are independent of the molecular details.
Specifically, $X(\tau_c)$ is a time-varying function whose temporal profile is almost identical to that of the pump pulse (see \Figs{fig1}a and \ref{fig2}a). $Y(\zeta_P)$ and $Z(\zeta_P)$ are enhancement factors which originate from the strong longitudinal electromagnetic fields generated by the pump and the probe VBs, respectively.
Particularly, the $Z$ factor increases drastically with the decrease of $\zeta_P$ (\Fig{fig2}b).
Moreover, the intensity of the VBMCD signal is inversely proportional to the quadratic of $\zeta_L$.   
Therefore, using a shorter and thinner sample cuvette is generally more favorable to achieve an enhanced VBMCD spectrum.
Nevertheless, in practice the sample cuvette cannot be too small, 
because too few molecules could make the luminous flux in the radial direction
too weak to detect or the signal-to-noise ratio too low.

To demonstrate the utility of the proposed VBMCD protocol,
we perform simulations for the MCD signals of the hydroxyl (OH) radical (Section~S2 in the SI).
\Figs{fig2}c and \ref{fig2}d depict the lineshape and intensity of the VBMCD spectrum,
which are compared in parallel with the conventional MCD spectrum 
[$\Delta \epsilon_{\rm PW}(\omega)$] probed by PWs
(see Equation~S12 in the SI for more details).
The Faraday $A$, $B$ and $C$ terms of the VBMCD spectrum agree closely with their counterparts of the conventional MCD spectrum.
This confirms that the VBMCD protocol reproduces all the important spectroscopic information rendered by conventional MCD.

By definition, it is clear that the signal acquisition time does not affect
the magnitude of absorption or MCD spectra, but does affect the signal-to-noise ratio in an experiment.
Hence, in order to compare the VBMCD and conventional MCD spectra on an equal basis
and to accentuate the importance of the enhancement factors,
in our simulations we assume that all measurements are performed with sufficiently high signal-to-noise ratio with long enough acquisition time
and that the VBs and PWs are equally affected by the optical elements (not shown in \Fig{fig1}).
In \Fig{fig2}d, the localized magnetic field $\mathcal B_z$ generated by the pump pulse has a peak value of $0.74\,{\rm T}$, about $10$ times weaker than the static magnetic field $B_z$ ($\sim 7.4\,{\rm T}$) adopted for the conventional MCD.
However, it is remarkable to see that the intensity of $\Delta \epsilon_\rho(\omega)$ is $500$ times stronger than the conventional MCD $\Delta \epsilon_{\rm PW}(\omega)$.
Although the intensity of both VBMCD signal and conventional MCD signal
scales linearly with the magnetic field (see \Equation{eq_2} and Section~S1 in the SI),  
the radially detected VBMCD signal can be considerably enhanced by properly adjusting the enhancement factors in \Equation{eq_7}, 
which is not possible for the conventional MCD.
%
%
These results verify that the enhancement factors associated with the radial
detection scheme cause the significant enhancement of the VBMCD signal in \Fig{fig2}d.
Moreover, a pulsed VB is superior to a continuous-wave VB as the pump, since the former can yield a stronger transient magnetic field by consuming the same amount of power.
It is possible to further enhance the VBMCD signal by tailoring the VBs.
For instance, by aiming the AP VB to a metallic circular aperture,\cite{blanco2018ultraintense}
the amplitude of $\mathcal B_z$ can be further amplified by $\sim 3.8$ times, reaching a peak value of $2.8\,{\rm T}$ 
(see Section~S1 in the SI for more details).
Specifically, by setting the probe beam waist, and the radius and length of the sample cuvette to
$w_0=1 \,{\rm mm}$, $P=0.1 \,{\rm mm}$ and $L=0.03 \,{\rm mm}$,\cite{berova2011comprehensive} respectively,
the VBMCD signal will reach an intensity about $4$ orders of magnitude stronger than the conventional MCD.
%

%
%

\begin{figure}[htbp]
  \centering
  \includegraphics[width=0.7\columnwidth]{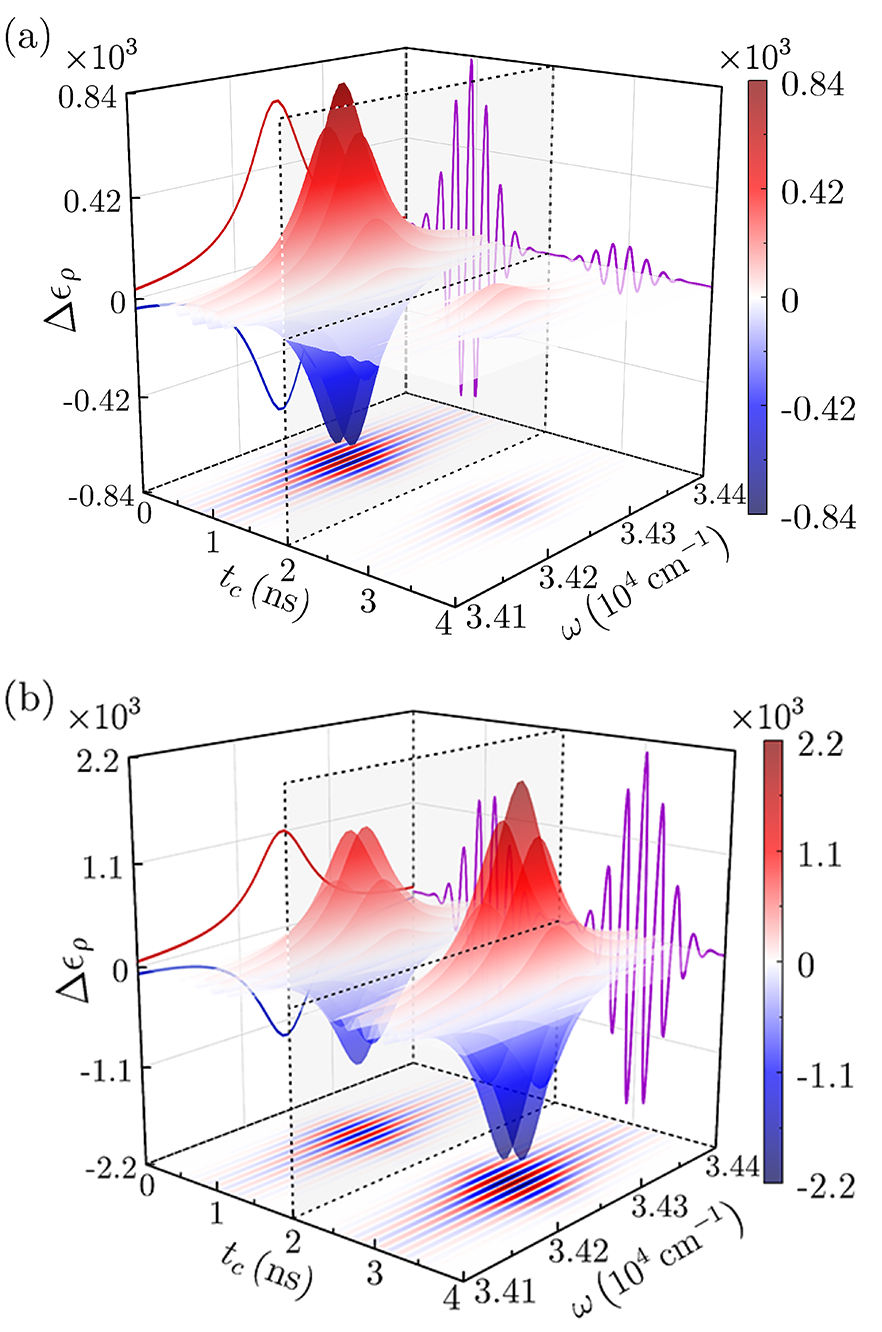}
  \caption{
  Simulated time-resolved VBMCD spectra of OH radicals, $\Delta \epsilon_\rho(\omega,\tau_c)$, in two scenarios characterized by the time-varying molecular concentration:
(a) $n(t)=n_0 e^{-rt}$ and (b) $n(t)=n_0 (1-e^{-rt})$, with the rate constant $r=1 \, {\rm ns}^{-1}$. Both scenarios adopt two temporally separated pump pulses centered at $T_c=1\,{\rm ns}$ and $3 \,{\rm ns}$, and the vertical plane at $t=2 \,{\rm ns}$ (indicated by dashed lines) marks the center of interval between the two pump pulses. Note that the time delay has been converted to the probe central time, i.e., $t_c=T_c+\tau_c$. The other parameters adopted for the simulations are listed in Section~S3 in the SI.
 }  \label{fig3}
\end{figure}

Pulsed VBs enable a magnet-free experimental protocol and achieve a substantial enhancement of MCD signals, and further allow for the detection of real-time dynamics of molecular systems by tuning the time delay between the pump and probe pulses. In the present time-resolved VBMCD protocol, the pump pulses do not trigger any dynamic change of molecules. Instead, they impose weak perturbation on the molecular electronic states by creating Zeeman splitting, so that the molecules become discernible to the subsequent probe pulses.

As a demonstration, we consider two scenarios, in which OH radicals are consumed or produced at the nanosecond time scale, respectively. The corresponding time-resolved VBMCD spectra, $\Delta \epsilon_\rho(\omega,\tau_c)$, are displayed in \Fig{fig3}a and \ref{fig3}b. In each scenario, two temporally separated pump pulses are applied, and the relative intensities of the MCD signal detected at the probe central time $t_c$ directly reflect the change in molecular concentration over time, since $\Delta \epsilon_\rho \propto n(t)$. The signal changes its sign periodically with $t_c$, due to the $X(\tau_c)$ factor in \Equation{eq_7}, and the signal intensity reaches its maximum when the pump-probe time delay is zero, because the corresponding pump magnetic field is the strongest.
In \Fig{fig3}a, the signal intensity is much reduced at the second pump pulse, which reveals a rapid drop in concentration;
whereas in \Fig{fig3}b, the signal is somewhat enhanced at the second pump pulse,
indicating a mild increase in concentration.
\begin{figure}[h]
\centering
  \includegraphics[width=0.7\columnwidth]{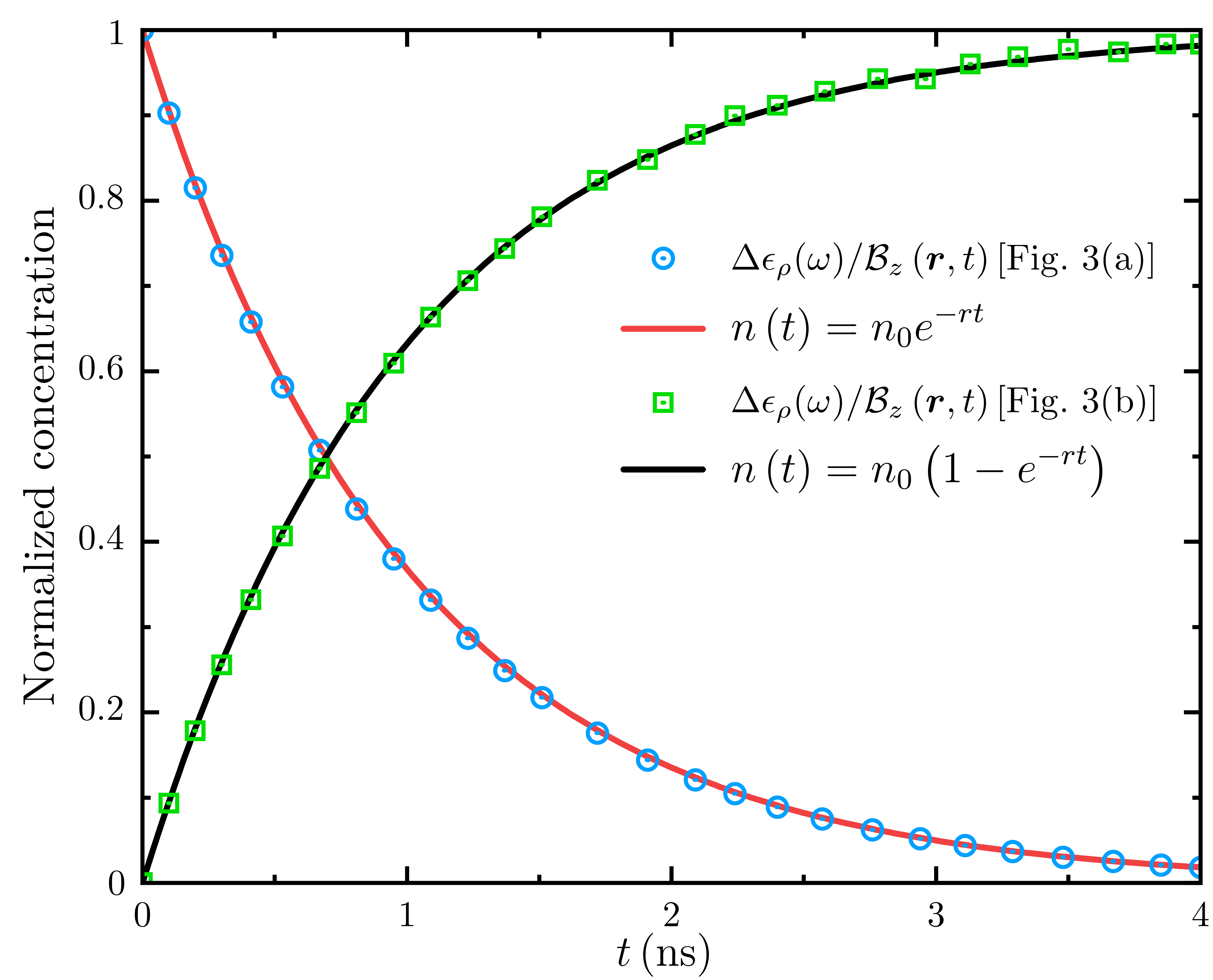}
  \caption{Real-time variations of OH radical concentration calculated by $\Delta \epsilon_\rho(\omega,\tau_c)/\mathcal{B}_z(\bm r,t)$ 
  for the two scenarios explored in \Fig{fig3}.
The actual variations of $n(t)$ are also plotted for comparison.
The frequency $\omega$ is fixed at $\omega=34250\,\,{\rm cm}^{-1}$.
The other parameters adopted for the simulations are listed in Section~S3 in the SI.}
\label{fig4}
\end{figure}

As inferred from \Equations{eq_2} and \ref{eq_7},
the time dependence of the VBMCD spectra is governed
by the molecular concentration $n(t)$ and the pump magnetic field
$\mathcal{B}_z(\bm r,t)$, and the parameters for the latter are
already known and can be precisely controlled when conducting an experiment.
Therefore, by performing the calculation for
$\Delta \epsilon_\rho(\omega,\tau_c)/\mathcal{B}_z(\bm r,t)$,
we can extract the time-dependent molecular concentration
from the measured VBMCD spectra.
As exemplified in \Fig{fig4}, the calculated data
for the OH radical consumption and production processes (\Fig{fig3})
agree closely with the actual concentrations.
Hence, the time-resolved VBMCD signal
provides a route to extract information about the real-time variation of the
molecular concentration.

Note that, in addition to the dynamic variation of molecular concentration at the nanosecond time scale, the proposed protocol may be conveniently generalized to the investigation of geometric and electronic dynamics of magnetic molecules or materials at the picosecond time scale, e.g., the ultrafast magnetization of ferromagnets triggered by a certain driving source.\cite{higley2016femtosecond,leveille2022ultrafast}
This can be achieved by adopting a picosecond pump AP VB in our protocol. Therefore, the time-resolved VBMCD protocol offers a useful spectroscopic tool for monitoring ultrafast dynamic process of molecules.
%

%
To conclude, we have theoretically designed an efficient protocol for realizing the significantly enhanced magnet-free MCD measurement, where the strong longitudinal magnetic field of the nanosecond GHz AP VB induces the Zeeman splitting in the molecules, and the femtosecond UV-Vis LCPL/RCPL enables the enhancement of the MCD signal. Furthermore, the time-resolved signal obtained by varying the pump-probe time delay promises to unravel transient molecular dynamics on the nanosecond time scale. The proposed protocol is currently feasible and may boost the investigation of ultrafast dynamic processes in magnetic molecules and materials.
%
%
%


%

\subsection*{Supporting Information Available}
All other data supporting this work, including derivation of the formulas of the VBMCD spectrum, quantum chemistry calculations for the VBMCD spectrum of the OH radicals, and parameters adopted for the simulations illustrated in \Figs{fig2}, \ref{fig3} and \ref{fig4}, are
provided in the Supporting Information. 

\section*{Acknowledgements}
J.C., L.Y., D.H., and X.Z. acknowledge the support from the National Natural Science Foundation of China (Grant Nos. 21973086 and 22203083), 
the Fundamental Research Funds for the Central Universities (Grant No.~WK2060000018), 
and the Ministry of Education of China (111 Project Grant No. B18051). S.M. was supported by the National Science Foundation (NSF) under Grant CHE-1953045, and the U.S. Department of Energy (DOE), Office of Science, Office of Basic Energy Sciences under Award Number DE-SC0022134. Computational resources are provided by the Supercomputing Center of University of Science and Technology of China.

\bibliography{ref}

%


\end{document}